\documentstyle[aps,prl,twocolumn,epsf]{revtex}
\begin{document}
\def\B.#1{{\bbox{#1}}}
\title{{\rm PHYSICAL REVIEW LETTERS \hfill submitted  }\\
Non-Perturbative Zero Modes in the Kraichnan Model for Turbulent 
Advection}
  \author {Omri Gat, Victor S. L'vov, Evgenii Podivilov and
Itamar  Procaccia}
  \address{Department of~~Chemical Physics,
 The Weizmann Institute of Science,
  Rehovot 76100, Israel}
 \maketitle
 \begin{abstract}
   The anomalous scaling behavior of the $n$-th order correlation
   functions ${\cal F}_n$ of the Kraichnan model of turbulent passive
   scalar advection is believed to be dominated by the homogeneous
   solutions (zero-modes) of the Kraichnan equation $\hat{\cal
     B}_n{\cal F}_n=0$. Previous analysis found zero-modes in
   perturbation theory in a small parameter. We present
   non-perturbative analysis of the simplest (non-trivial) case of
   $n=3$ and compare the results with the perturbative predictions.
\end{abstract}

The Kraichnan model of turbulent passive scalar advection \cite{68Kra}
has attracted enormous attention recently,
\cite{94Kra,95GK,95CFKL,95KYC,96FGLP} being the first non-trivial
model of turbulent statistics in which the phenomenon of multi-scaling
seems understandable by analytic methods. The model is for a scalar
field $T(\B.r,t)$ which satisfies the equation of motion \FL
\begin{equation}
{\partial T(\B.r,t) \over \partial t}+
\B.u(\B.r,t)\cdot\nabla T(\B.r,t)=\kappa \nabla^2
T(\B.r,t)+\xi(\B.r,t). \label{eq}
\end{equation}
Here $\xi(\B.r,t)$ is a Gaussian white random force, $\kappa$ is the
diffusivity and the driving field $\B.u(\B.r,t)$ is chosen to have
Gaussian statistics, and to be ``fastly varying" in the sense that its
time correlation function is proportional to $\delta(t)$. The
statistical quantities that one is interested in are the many point
correlation functions
\begin{equation}
{\cal F}_{2n}({\bf r}_1,{\bf r}_2,...,{\bf r}_{2n})
 \equiv \langle \!\langle T(\B.r_1,t)
T(\B.r_2,t)\dots T(\B.r_{2n},t)\rangle \! \rangle ,
\end{equation}
where double pointed brackets denote an ensemble average with respect
to a stationary statistics of the forcing {\em and } the statistics of
the
velocity field. One of Kraichnan's major results \cite{94Kra} is an
exact differential equation for this correlation function,
\begin{equation}
\big[- \kappa \sum_{\alpha} \nabla^2_\alpha + \hat {\cal B}_{2n}
 \big] {\cal F}_{2n}({\bf r}_1,{\bf r}_2,...,{\bf r}_{2n})
 = {\rm RHS} . \label{difeq}
\end{equation}
The operator $\hat {\cal B}_{2n}\equiv \sum_{\alpha>\beta}^{2n}
\hat{\cal B} _{\alpha\beta}$, and $\hat{\cal B} _{\alpha\beta}$ are
defined by
\begin{equation}
\hat{\cal B}_{\alpha \beta}\equiv \hat {\cal B}
({\bf r}_\alpha,{\bf r}_\beta)
         = h_{ij}({\bf r}_\alpha -{\bf r}_\beta) \partial^2 /
        \partial r_{\alpha,i} \partial r_{\beta,j}  \ ,
\end{equation}
where the ``eddy-diffusivity" tensor $ h_{ij}(\B.R)$ is given by
$$
h_{ij}({\B.R}) = h(R) [(\zeta_h +d-1) \delta_{ij} - \zeta_h R_i
R_j/R^2], $$
and $h(R) = H (R/{\cal L})^{\zeta_h}$, $0\le \zeta_h\le
2$.  Here ${\cal L}$ is some characteristic outer scale of the driving
velocity field.  The scaling properties of the scalar depend
sensitively on the scaling exponent $\zeta_h$ that characterizes the
$R$ dependence of $h_{ij}({\B.R})$ and that can take values in the
interval $[0,2]$. Finally, the RHS in Eq.(\ref{difeq}) is known
explicitly, but is not needed here. The reason is that it was argued
that the solutions of this equation for $n>1$ are dominated by the
homogeneous solutions (``zero-modes"), in the sense that deep in the
inertial interval the inhomogeneous solutions are negligible compared
to the homogeneous one. Also, it was claimed that in the inertial
interval one can neglect the Laplacian operators in Eq.(\ref{difeq}),
and remain with the simpler homogeneous equation $\hat {\cal B}_{2n}
{\cal F}_{2n}=0$.

Having exact differential equations for ${\cal F}_{2n}$ allowed
Kraichnan to announce a mechanism for anomalous scaling \cite{94Kra}.
Assuming that the physical solutions are scale invariant one needs to
examine the scaling (or homogeneity) exponent $\zeta_{2n}$ of ${\cal
  F}_{2n}$ which is defined by ${\cal F}_{2n}(\lambda \B.r_1,\lambda
\B.r_2\dots \lambda \B.r_{2n}) =\lambda^{\zeta_{2n}} {\cal
  F}_{2n}(\B.r_1, \B.r_2\dots \B.r_{2n})$ if such a solution exists.
One expects it to exist in the inertial range, i.e. all the
separations $r_{ij}$ satisfy $\eta\ll r_{ij}\ll L$ where $\eta$ and
$L$ are the inner and outer scales respectively. It is known
\cite{68Kra} that for ${\cal F}_2$ such a solution exists with
$\zeta_2=2-\zeta_h$.  If one solves for these exponent for $n>1$, one
can understand, at least in this simple model, what are the mechanisms
for deviations from the predictions of dimensional analysis, with
possible insight also for the Navier-Stokes problem.  In searching
methods for computing these important exponents, there emerged two
basic strategies. One strategy considered the differential equation in
the ``fully unfused" regime in which all the separations between the
coordinates are in the inertial range.  In this case even in the
simplest case of $n=2$ the function ${\cal F}_{4}$ depends on six
independent variables (for dimensions $d>2$), and one faces a
formidable analytic difficulty for exact solutions. Accordingly,
several groups considered perturbative solutions in some small
parameter, like $\zeta_h$ \cite{95GK} or the inverse dimensionality
$1/d$ \cite{95CFKL}.  The rationale for this approach is that at
$\zeta_h=0$ and $d\to \infty$ one expects ``simple scaling" with
$\zeta_{2n}=n \zeta_2$. The exponent $\zeta_4$, and later also the set
$\zeta_{2n}$, were computed as a function of $\zeta_h$ near these
simple scaling limits. Another strategy considered the differential
equation in the ``fully fused" regime, in which the correlation
function degenerates to the structure function $S_{2n}(R)=\langle\!
\langle [T(\B.r+\B.R)- T(\B.r)]^{2n}\rangle\! \rangle$. In this approach
there is an enormous simplification in having only one variable, but
one loses information in the process of fusion.  The lost information
was supplemented \cite{94Kra} by a yet underived conjecture about the
properties of conditional averages, leading at the end to a close-form
calculation of the exponents $\zeta_{2n}$ for arbitrary dimension and
values of $\zeta_h$. The results of the two strategies are not in
agreement.  Even though numerical simulations \cite{95KYC} and also
experiments \cite{96CLP,96CLPP} lend support to the assumption used in
the second strategy and to its computed values of $\zeta_{2n}$, there
remains an important mystery as to why the two approaches reach such
different conclusions. The aim of this Letter is to explore
non-perturbative calculations of the zero modes and their exponents,
to shed further light on this issue.

Our strategy is to solve exactly, eigenfunctions included, the
homogeneous equation satisfied by the 3'rd order correlation function
${\cal F}_3(\B.r_1,\B.r_2,\B.r_3)$.  Note that in Kraichnan's model
all the odd-order correlation functions ${\cal F}_{2n+1}$ are zero
because of symmetry under the transformation $T\to -T$. This symmetry
disappears for example \cite{example} if the random force $\xi(\B.r,t)$
is not Gaussian (but $\delta$-correlated in time), and in particular
if it has a non-zero third order correlation \FL
\begin{equation}
{\cal D}_3(\B.r_1,\B.r_2,\B.r_3)\equiv \!\int\!
 dt_1 dt_2\langle \xi(\B.r_1,t_1)
\xi(\B.r_2,t_2)\xi(\B.r_3,0) \rangle. \label{D3}
\end{equation}
With such a forcing the third order correlator is non-zero, and it
satisfies the equation
\begin{equation}
\hat {\cal B}_3 {\cal F}_3(\B.r_1,\B.r_2,\B.r_3) =
 {\cal D}_3\,, \quad
\hat {\cal B}_3\equiv\hat {\cal B}_{12}+
\hat {\cal B}_{13}+\hat {\cal B}_{23}\ . \label{sum}
\end{equation}
This equation pertains to the inertial interval and accordingly we
neglected the Laplacian operators. We also denoted ${\cal D}_3
=\lim_{\B.r_{\alpha\beta} \to 0} {\cal D}_3(\B.r_1,\B.r_2,\B.r_3)$.
The solution of this equation is a sum of inhomogeneous and homogeneous
contributions, and below we study the latter. We will focus on scale
invariant homogeneous solutions which satisfy ${\cal
  F}_3(\lambda\B.r_1,\lambda\B.r_2,\lambda\B.r_3)=\lambda^{\zeta_3}
{\cal F}_3(\B.r_1,\B.r_2,\B.r_3)$. We refer to these as the ``zero
modes in the scale invariant sector". We note that the scaling
exponent of the {\em inhomogeneous} scale invariant contribution can
be read directly from power counting in Eq.(\ref{sum}) (leading to
$\zeta_3=\zeta_2$). Any different scaling exponent can arise only from
homogeneous solutions that do not need to balance the constant RHS. In
addition, note that scale-invariant zero-modes arise not only due to
the omission of the diffusive terms from Eq. (\ref{sum}), but also as
a result of the omission of the boundary conditions for large
separation (at the outer scale $L$).  The smooth connection to
either small or large scales must
ruin scale invariance. The scale invariant solutions of Eq.(\ref{sum})
live in a projective space whose dimension is lowered by unity
compared to the most general form; These solutions do not depend
on three separations but rather on two dimensionless variables that 
are identified below. It will be demonstrated how boundary conditions
arise 
in this space for which
the operator $\hat {\cal B}_3$ is neither positive nor self-adjoint.

Equation (\ref{sum}) is also invariant under the action of the $d$
  dimensional rotation group SO($d$), and under permutations of the
three
  coordinates. Here we seek solutions in the scalar representation of
  SO($d$), where the solution depends on the 3 separations $r_{12}$,
  $r_{23}$ and $r_{31}$ only. We transform coordinates to the
  variables $x_1=|\B.r_2-\B.r_3|^2$, $x_2=|\B.r_3-\B.r_1|^2$,
  $x_3=|\B.r_1-\B.r_2|^2$.  The triangle inequalities in the original
  space are equivalent to the condition
\begin{equation}
\label{ineq}
2(x_1x_2+x_2x_3+x_3x_1)\ge x_1^2+x_2^2+x_3^2.
\end{equation}
The advantage of the new coordinates is that the inequality
(\ref{ineq}) describes a circular cone in the $x_1,\,x_2,\,x_3$ space
whose axis is the line $x_1=x_2=x_3$ and whose circular cross section
is tangent to the planes $x_1=0$, $x_2=0$ and $x_3=0$.
This cone can be parameterized by three new coordinates
$s,\,\rho,\,\phi$:
\begin{eqnarray}
&&x_n=s\{1-\rho\cos[\phi+(2\pi/3)n]\},\nonumber\\
&&0\le s<\infty,\quad 0\le \rho\le1,\quad 0\le\phi\le2\pi.
\end{eqnarray}
The $s$ coordinate measures the overall scale of the triangle
defined by the original $\B.r_i$ coordinates, and configurations of
constant $\rho$ and $\phi$ correspond to similar triangles. The $\rho$
coordinate describes the deviation of the triangle from the
equilateral configuration ($\rho=0$) up to the physical limit of three
collinear points attained when $\rho=1$; $\phi$ does not have a simple
geometric meaning. 

The transformation of the linear operator $\hat {\cal B}_3$ to the new
coordinates is straightforward, and produces a second order linear
partial differential operator in the $s,\,\rho,\,\phi$ variables (the
full form of the operator is long and will not be given here). The
scale invariant solution take on the form $s^{\zeta_3/2}f(\rho,\phi)$,
and the transformed operator applied to this form gives an equation
for $f(t,\phi)$
\begin{eqnarray}
&&\hat  B_3(\zeta_3) f(\rho,\phi)=[a(\rho,\phi)
\partial_\rho^2+b(\rho,\phi)\partial_\phi^2+
c(\rho,\phi)\partial_\rho\partial_\phi
 \label{rdeq}\\
&&+u(\rho,\phi,\zeta_3)\partial_\rho+v(\rho,\phi,
\zeta_3)\partial_\phi+w(\rho,\phi,\zeta_3)]
f(\rho,\phi)=0 \ .\nonumber
\end{eqnarray}
The new operator $\hat B_3$ depends on $\zeta_3$ as a parameter and it
acts on the unit circle described by the polar $\rho,\phi$
coordinates. The circle represents the projective space of the
physical cone described above.

The discrete permutation symmetry of the original Eq.(\ref{sum})
results in a symmetry of Eq.(\ref{rdeq}) with respect to the 6 element
group generated by the transformation $\phi\rightarrow\phi+2\pi/3$
(cyclic permutation of the coordinates in the physical space) and
$\phi\rightarrow-\phi$ (exchange of coordinates). This symmetry
extends to a full $U(1)$ symmetry in the two marginal cases of
$\zeta_h=0$ and $\zeta_h=2$ for which all the coefficients in
(\ref{rdeq}) become $\phi$-independent. The coefficients in
(\ref{rdeq}) all have a similar structure, and for example
$a(\rho,\phi)$ reads
$$
a(\rho,\phi)=\sum_n[1-\rho \cos(\phi+\case{2}{3}\pi n)]
^{(\zeta_h-2)/ 2}\tilde a(\rho,\phi+\case{2}{3}\pi n) \ ,
$$
where $\tilde a(\rho,\phi)$ is a low order polynomial in $\rho$,
$\cos\phi$ and $\sin\phi$ which vanishes at $\rho=1,
\phi=0$. We see that the coefficients are analytic
everywhere
\begin{figure}
\epsfxsize=8truecm
\label{Fig1}
\caption{ The scaling exponent $\zeta_3$ as a functions of $\zeta_h$
found as the loci of zeros of the determinant of the matrix $B_3$, for
$d=2$.} 
\end{figure}

\noindent
on the circle except at the three points $\rho=1$, $\phi=2\pi n/3$
where $n=0,1,2$. These points correspond to the fusion of one pair of
coordinates, and the coefficients exhibit a branch point singularity
there.  This singularity leads to a nontrivial asymptotic behavior of
the solutions which had been described before in terms of the fusion
rules \cite{96FGLP,96LP}. Note that for $\zeta_h=2$ the singularity
disappears trivially. For $\zeta_h=0$ there is also no singularity
since $\tilde a$ exactly compensates for the inverse power.

The boundary conditions follow naturally when one realizes that $\hat
B_3$ is elliptic for points strictly inside the physical circle. This
is a consequence of the ellipticity of the original operator $\hat
{\cal B}_3$. On the other hand $\hat B_3$ becomes singular on the
boundary $\rho=1$, where the coefficients $a(\rho,\phi)$ and
$c(\rho,\phi)$ vanish. This singularity reflects the fact that this is
the boundary of the physical region. It follows that $\hat B_3$
restricted to the boundary becomes a relation between the function
$f(\rho=1,\phi)\equiv g(\phi)$ and its normal derivative
$\partial_\rho f(\rho,\phi)\vert_{\rho=1}\equiv h(\phi)$. The relation
is $bg''+uh+vg'+wg=0$. Solutions of Eq.(\ref{rdeq}) which do not
satisfy this boundary condition are singular, with infinite $\rho$
derivatives at $\rho=1$. Such solutions are not physical since
they involve infinite correlations between the dissipation (second
derivative of the field) and the field itself when the geometry
becomes collinear, but without fusion.

Having a homogeneous equation with homogeneous boundary conditions we
realize that non-trivial solutions are available only when $\det(\hat
B_3)=0$. This determinant depends parametrically on $\zeta_3$.  Since
the operator is defined on a compact domain we expect the determinant
to vanish only for discrete values of $\zeta_3$ for any given values
of $\zeta_h$ and the dimensionality $d$. We know that there always
exists a trivial constant solution associated with $\zeta_3=0$. Our 
aim is to find the lowest lying
positive real solutions $\zeta_3$ for which the determinant vanishes.
\begin{figure}
\epsfxsize=8truecm
\label{Fig2}
\caption{ Same as Fig.1, but for $d=3$.}
\end{figure}
\begin{figure}
\epsfxsize=8truecm
\label{Fig3}
\caption{ Same as Fig.1, but for $d=4$. }
\end{figure}
\noindent
We approach the problem numerically by discretizing the operator $\hat
B_3$ including the boundary conditions, and solving the analogous
problem for the discretized operator. Using the symmetry of the
problem we restricted the domain to one sixth of the circle, and
defined a nine-point finite difference scheme for the evaluation of
the second order derivatives. The discretized boundary conditions at
$\rho=1$ were achieved with the same scheme. The symmetry implies that
the new boundary conditions on the lines $\phi=0,\pi/3$ are simple
Neuman boundary conditions $\partial_\phi f(\rho,\phi)=0$. After
discretization the problem transforms to a matrix eigenvalue problem
$B_3 \B.\Psi=0$, where $B_3$ is a large sparse matrix, whose rank
depends on the mesh of the discretization, and $\B.\Psi$ is the
discretized $f$.  We used NAG's sparse Gaussian elimination routines
to find the zeros of $\det(B_3)$, and determined the values of
$\zeta_3$ for these zeros as a function of $\zeta_h$.  The results of
this procedure for space dimensions $d=2,3,4$ are presented in Figs.
1,2, and 3.

The various branches shown in Figs. 1-3 can be organized on the 
basis of the perturbation theory of the type proposed in
\cite{95GK} near $\zeta_h=0$. We performed that 
type of analysis and found that at $\zeta_h=0$ the allowed values
of $\zeta_3$ are organized in two sets,
\begin{eqnarray}
\zeta_3^+(m,n)&=&2(3m+2n)\,, \nonumber \\
 \zeta_3^-(m,n)&=&-2(d-1+3m+2n) \,, \label{pert}
\end{eqnarray}
where $n$ and $m$ are any non-negative integer. The lowest lying
positive values are $4,6,8$ etc, whereas for $d=2$ the highest
negative value is $-2$.  We see that the nonperturbative solution
displays in all dimensions a branch (dashed line) which begins at
$\zeta_h=0,\zeta_3=4$ and ends at $\zeta_h=2,\zeta_3=0$.  This branch
is identical to the lowest lying positive branch predicted by the
perturbation theory. We computed the slope of this branch near
$\zeta_h=0$ in perturbation theory, and found that it is
$2(2-d)/(d-1)$, in agreement with the numerics. Also the slopes of the
other branches that begin at $\zeta_h=0$ were obtained perturbatively
and found to agree with the numerics.  The negative branch (shown only
for $d=2$) never rises above its perturbative limit and is not
relevant for the scaling behaviour at any value of $\zeta_h$. Note
also that the point $\zeta_h=2, \zeta_3=0$ appears to be an
accumulation point of many branches, and we are not confident that all
the branches there were identified by our finite discretization
scheme. This raises a worry about the availability of a smooth
perturbative theory around $\zeta_h=2$. At least we expect such a
perturbation theory to be very singular. Preliminary analytical work
indicates that all the branches join the point $\zeta_h=2, \zeta_3=0$
with an infinite slope.

The results of our nonperturbative approach lend support to the
validity of the perturbative calculations of the zero modes of $\hat
B_4$. The disagreement between the scaling exponents $\zeta_4$ and the
higher order exponents $\zeta_n$ computed via the perturbative
approach and the predictions of the other approach based on the fully
fused theory cannot be ascribed to a formal failure of the
perturbation theory. There are therefore a few possibilities
that have to be sorted out by further research:\\
(i) The crucial assumption that goes to the fully fused approach,
which is the linearity of the conditional average of the Laplacian
of the scalar, is wrong. \\
(ii) The computation of the zero modes which is achieved by discarding
the viscous terms in $\hat B_n$ is irrelevant for the physical
solution. It is not impossible that the diffusive term act as a
singular perturbation on some of the scale invariant modes. In fact,
the operator $\hat B_n$ with the viscous term is positive definite and
it has no zero modes. That this is a possibility is underlined by
recent calculations of a shell model of the Kraichnan model
\cite{96WB}, in which it was shown that the addition of any minute
diffusivity changes
the nature of the zero modes qualitatively.\\
(iii) Lastly, and maybe most interestingly, it is possible that the
physical solution is not scale invariant \cite{Bob}. In other words,
it is possible that ${\cal F}_3(\B.r_1, \B.r_2,\B.r_3)$ is not a
homogeneous functions with a fixed homogeneity exponent $\zeta_3$, but
rather (for example), that $\zeta_3$ depends on the ratios of the
separations (or, in other words, the geometry of the triangle defined
by the coordinates). If this were also the case for even correlation
functions ${\cal F}_{2n}$, this would open an exciting route for
further research to understand how non-scale invariant correlation
functions turn, upon fusion, to scale invariant structure functions.

In light of the numerical results of ref.\cite{95KYC} and the 
experimental results displayed in \cite{96CLP,96CLPP} we tend to doubt
option (i). If we were to guess at this point we would opt for 
possibility (ii). More work however is needed to clarify this
important issue beyond doubt.

{\bf Acknowledgments}: We thank Bob Kraichnan for useful suggestions
and J-P. Eckmann and Z. Olami for discussions. This
work was supported in part by the US-Israel BSF, the German-Israeli
Foundation, the Minerva Center for Nonlinear Physics and the 
Naftali and Anna Backenroth-Bronicki Fund for Research in Chaos and
Complexity.



\begin{references}
\bibitem{example} There are other possibilities that make ${\cal F}_3$
  relevant. One is putting on a fixed gradient, see \cite{95SS}, and
  another is to consider the fusion of three points in a higher order
  correlations, see \cite{96FGLP}.
\end{references}
\end{document}